# Evidence for an Anti-polar Phase
# In Normal and Superconductivity States in all HTSC

T. GUERFI[*]

*Department of physics. Faculty of Sciences.*
*M'hamed Bougara University, Boumerdes.*
*35000. Algeria*
*E-mail address:* tarek.guerfi @gmail.com

**Abstract**

It is strongly argued that high-temperature superconductors (HTSC) exhibit an anti-polar phase state with long range order in normal and superconducting states. This anti-polar phase is directly related to the onset of superconductivity in all high temperature superconductors and it is responsible for strong coupling and of two dimensionality aspect of HTSC, as it is described below.

Research on the crystal structures of high-temperature superconductors and structural changes with temperature is very helpful to understand the mechanism of the superconductivity. Therefore neutron and x-rays diffractions have been powerful methods in the study of high Tc superconductors. [1].

In the other hand a large number of experimental and theoretical investigations has indicated that two dimensionality aspects of the normal and superconducting state is one of the key factors in high temperature superconductivity [2, 3]. In Copper oxides (cuprates), it was found that the Ginzburg-Landau coherence length is less than the lattice spacing in the direction perpendicular to the CuO$_2$ planes (c-direction). Moreover, small buckling of these planes (on a scale of 1%) significantly reduces T$_c$ [4].

Superconductivity is of course a quantum phenomenon. However, classical arguments can shed important light into the fundamental physics of superconductors. In the sense of Bohr's correspondence principle, one can argue that the macroscopic quantum manifestations of superconductivity should be also understandable from a classical point of view.

In this letter, it is strongly argued that high-temperature superconductors exhibit an anti-polar phase with a long range order in both normal and superconducting states. The occurrence of this anti-polar sate phase is directly related to the superconducting condensation in all HTS and it is responsible for strong coupling and of two dimensionality aspect of HTSC in both normal and superconducting states

For one of the best studied compounds is YBa$_2$Cu$_3$O$_7$. A single phase specimen of superconducting **YBa$_2$Cu$_3$O$_{6.9}$** has been investigated by Schäfer et al [5], using high resolution neutron powder diffraction in the temperature range from 16 to 300 K. Based on the z atomic coordinate data result [8], we compute the dipole moment value upon the half of the YBCO unit cell using a simple formula:  $G^+ - G^- = \dfrac{\sum zq}{\sum q}$ where : $G^+$ and $G^-$ are the position of the centroid of the positive and the negative charges respectively upon the half unit cell along the c axis ( the polar axis); q the atomic oxidation.

The polarization is given by: $p = \left(G^+ - G^-\right) c \sum q$ :

where c the is c axis parameter of the unit cell.
We perform calculation for seven different temperatures 300K, 182K, 101K, 92 K, 82K and 16 K . The polarization has a value of **p (300 K) = 2.058(3) × 10$^{-29}$ Cm** along the c axis (the polar axis) at 300 K and **p(92 K) = 2.1670(8) × 10$^{-29}$ Cm** at 92 K which is the superconducting transition temperature.
The result is plotted in figure 1. We note a discontinuous change in the polarization near the superconducting transition temperature. Also, in their paper, W. Schäfer et al [8] showed structural anomalies of **YBa$_2$Cu$_3$O$_{6.9}$** at the superconducting transition temperature.
This provides the experimental evidence that YBCO material exhibit an anti-polar phase sate occurring both in normal and superconducting states. Figure 2 shows schematically the electric dipole moments distribution in the unit cell structure in case of YBCO compound.

Also using crystallographic data from [6] for **YBa$_2$Cu$_3$O$_{6.78}$** at room temperature, we find **p = 1.36 × 10$^{-29}$ Cm.**

For the compound **Hg$_{0.8}$Tl$_{0.2}$Ba$_2$Ca$_2$Cu$_3$O$_{8.09}$** using crystallographic data from [7] we find a value for polarization **p = 6.968(3) × 10$^{-29}$** Cm at temperature 200 K and **p = 6.967(0) × 10$^{-29}$ Cm** for the same compound at 170 K.

For the compound **TlBa$_2$CaCu$_2$O$_{6.5}$ (Tl-1212)**, using crystallography data from reference [8], we find **p = 2.250(6) × 10$^{-29}$ Cm** at room temperature.

For the 2212 type superconductor using the crystallography data from [9] the compound **Tl$_{1.3}$Hg$_{0.7}$Ba$_{1.5}$Sr$_{1.22}$Ca$_{0.28}$Cu$_2$O$_{7.54}$** gives a value of polarization **p = 9.5763(0) × 10$^{-29}$ Cm** at room temperature.

All high temperature superconductor materials exhibit an anti-polar phase in normal and superconducting states with a discontinuity of polarization near the superconducting transition. Structural instabilities and anomalies of HTSC near T$_c$, which can be considered as the crystallographic signature of the out of plane correlation between the electric dipole moments,(characterized by the discontinuity in the polarization) have been observed by many research groups using different experimental methods. For YBa$_2$Cu$_3$O$_x$ Srinivasan *et al* [10] claims that there is 0.4 °A anomalous jump in the *c* parameter (11.5 °A) of the lattice at the transition temperature whereas the variation in *a* and *b* parameters are negligible. Horn *et al* [11] has not observed a jump in the lattice parameters however, but he reports an anomaly in the orthorhombic splitting, *b-a*, near the superconducting transition whereas there is little or no anomaly in the unit cell volume. R.J. Cava et al [12] reported structural anomalies at the disappearance of superconductivity in YBCO material.

D. Cheng et al [13] reported structural anomalies occurring close to the superconducting phase transition in a nearly single phased Bi$_2$Sr$_2$Ca$_2$Cu$_3$O sample, also superconducting single crystals of Tl$_2$Ba$_2$CaCu$_2$O$_8$ and YBa$_2$Cu$_4$O$_8$ showed anomalous structural change in the vicinity of the critical temperature [14]. Structural anomaly just above Tc was also observed in (Hg$_{1-x}$Tl$_x$)Ba$_2$Ca$_2$Cu$_3$O$_{8-\delta}$ superconductor [15], Qinlun Xu *et al* [16] have reported structural changes in single phase samples of Bi(Pb)-Sr-Ca-Cu-O at 110 K. X.S. Wu et al [17] have also reported an anomaly in the structure of YBa$_{1.8}$La$_{0.2}$Cu$_3$O$_y$ near the superconducting transition and very recently similar to that in cuprates, lattice structure anomaly was reported in LaFeAsO$_{1-x}$F$_x$ at the onset superconducting transition temperature [18].

Dipole-dipole forces are short range and essentially anisotropic and their interaction can be either repulsive (positive dipole-dipole interaction energy) or attractive (negative dipole-dipole interaction energy). When a free charge carrier is introduced, which can be hole or electron depending on chemical doping, the current in normal state will result from two component the normal current carried by free charge carriers and the polarization current resulting from the In plane correlation between the dipole moments. Figure 3 gives a schematic picture of long range order in normal and superconducting states. In plane Correlations ( // ab) between eclectic dipoles moments are established in normal state while the Out of Plane ( $\perp$ ab) correlation take place only at the superconducting state.

In the superconducting state, the out of plane correlation, in addition to the in plane correlation, between the electric dipoles, take place. The parity symmetry is broken and a net Electric Dipole Moment (EDM) appears in the unit cell as consequence. It is noted that for single particle or a composite system the intrinsic EDM in non-degenerate state is also a signature of a parity and time-reversal broken symmetry.

In conclusion, it is shown that high-temperature superconductors exhibit an anti-polar state (long range order) both in normal and superconducting states. This phase is directly related both the metallic aspect in normal sate and to the superconductivity aspect in superconducting satet in all HTSC.

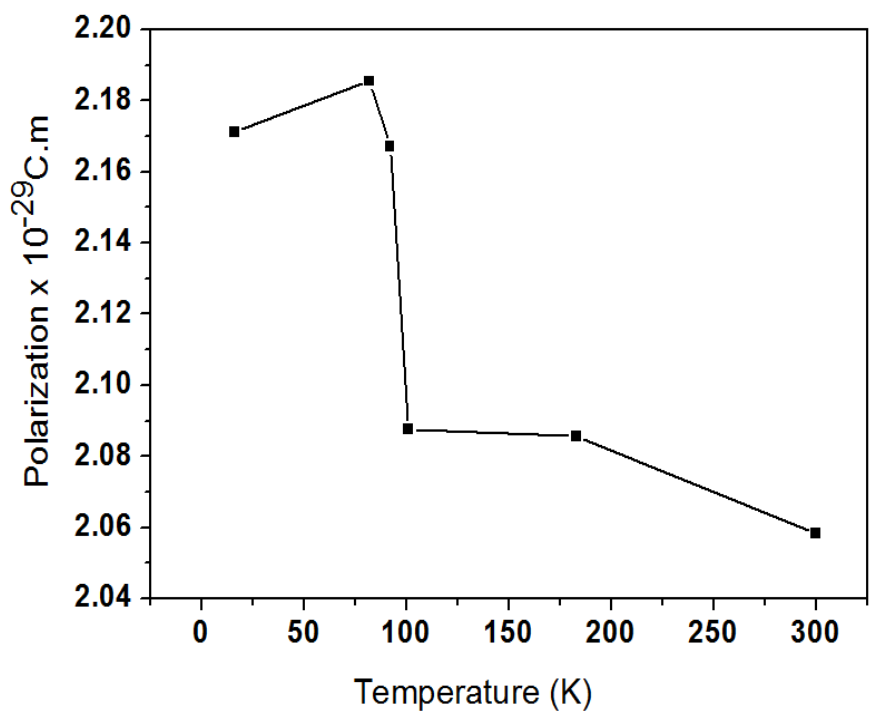

Figure 1. Temperature dependence of Polarization in YBa$_2$Cu$_3$O$_{6.9}$ unit cell.

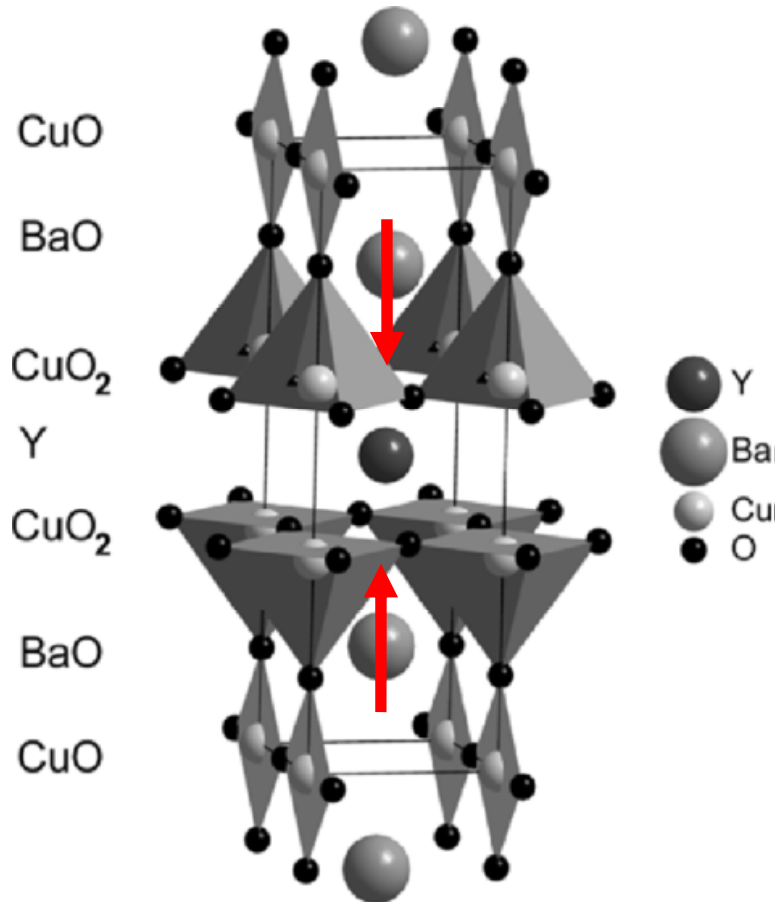

Figure 2. Schematic picture of anti-polar (long range order) phase occurring in $YBa_2Cu_3O_{6.9}$ unit cell in both normal and superconducting state. ( $\vec{p}$= 2.1607(5) .$10^{-29}$ Cm along the *c*-axis at 92K ) .

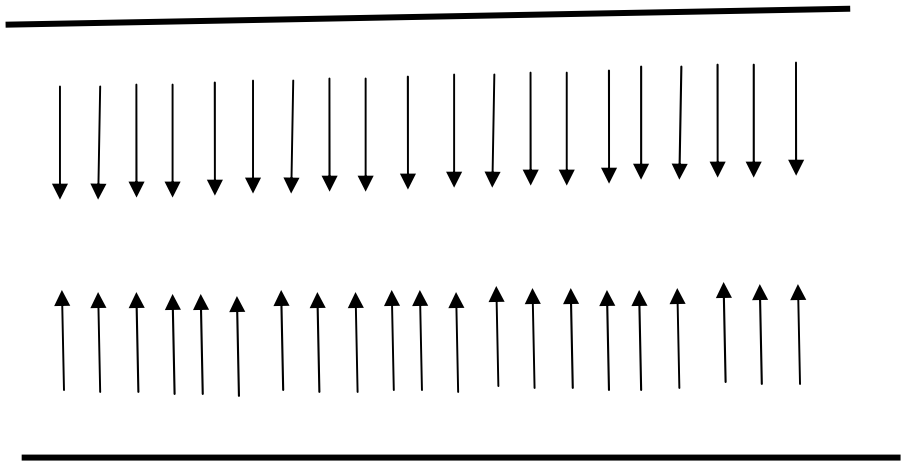

Figure.3. Schematic picture of long range order in normal and superconducting states. In plane Correlations ( // ab) between eclectic dipoles moments are established in normal state while the Out of Plane ( ⊥ ab) correlation take place only at the superconducting state.